\documentclass{article}

\usepackage{arxiv}

\usepackage[utf8]{inputenc} 
\usepackage[T1]{fontenc}    
\usepackage{url}            
\usepackage{booktabs}       
\usepackage{amsfonts}       
\usepackage{nicefrac}       
\usepackage{microtype}      
\usepackage{lipsum}
\usepackage{graphicx}
\usepackage{tabularx}
\usepackage{booktabs}
\usepackage{tabularx}
\usepackage{multirow}
\usepackage{caption}
\usepackage{graphicx}

\usepackage{subcaption}

\usepackage[sort&compress, numbers]{natbib}
\usepackage{hyperref}
\usepackage{amsmath}

\title{Machine Learning Surrogates for Optimizing Transportation Policies with Agent-Based Models \thanks{Extended abstract accepted for presentation at the 12\textsuperscript{th} Triennial Symposium on Transportation Analysis conference (TRISTAN XII), June 22-27, 2025, Okinawa, Japan}}

\author{
 Elena Natterer \thanks{Corresponding author} \\
  Chair of Traffic Engineering and Control \\
  Technical University of Munich, Germany \\
  \texttt{elena.natterer@tum.de} \\
   \And
 Roman Engelhardt \\
  Chair of Traffic Engineering and Control \\
  Technical University of Munich, Germany \\
  \texttt{roman.engelhardt@tum.de} \\
  \And
  Sebastian Hörl \\
  IRT System X, Paris, France\\
  \texttt{sebastian.horl@irt-systemx.fr} \\
    \And
 Klaus Bogenberger \\
  Chair of Traffic Engineering and Control \\
  Technical University of Munich, Germany \\
  \texttt{klaus.bogenberger@tum.de} \\
}

\begin{document}
\maketitle

\section{INTRODUCTION}

Urban growth intensifies traffic and pollution, necessitating tailored traffic management strategies such as congestion charges and repurposing car space, to reduce car usage during peak demand periods. 
For these measures to meet their intended goals and also achieve public acceptance, they must be designed carefully.


Agent-based simulations are often employed to assess how potential policy measures impact the broader transportation system. For instance, \cite{ben-dorSimulationbasedPolicyEvaluation2024} used them to evaluate congestion charges and parking fees aimed at reducing car traffic and alleviating congestion, while \cite{luSimulationbasedPolicyAnalysis2023} examined speed limit impacts on road safety, traffic flow efficiency, and environmental factors.

A major limitation of agent-based models is their extensive runtime, which restricts the number of scenarios that can be tested and thus hinders the ability to identify optimal policy setups. Techniques such as Bayesian Optimization have been proposed to reduce the number of required simulations, enabling more efficient solution space exploration (e.g.,~\cite{Dandl.2021,Huo.2023}).


In contrast, if machine learning (ML) models can replicate essential outcomes of these agent-based models, they could serve as fast-executing surrogates, enabling large-scale simulation-based optimization.
Previous research, such as \cite{RoadTrafficCanBePredictedByMLLikeMicroscopicModels}, demonstrated that ML models can predict road traffic as effectively as complex microscopic simulations. Graph Neural Networks (GNNs) are particularly promising for transportation forecasting due to their ability to incorporate spatial dependencies within urban networks directly into the neural network structure. \cite{jiangGraphNeuralNetwork2022} provided a comprehensive review of GNN-based traffic forecasting studies, highlighting available resources and research challenges.

This study aims to evaluate the feasibility of using Graph Neural Networks as surrogates for agent-based models in transport planning, with the long-term goal of enabling simulation-based optimization through agent-based simulations. To the authors’ knowledge, only \cite{Narayanan_2024} has provided a proof-of-concept, replicating the 4-step model with GNNs on networks of 15 to 80 nodes. Our study, by contrast, trains a GNN using MATSim simulation data for a Paris case study with 30,000 nodes. The results indicate that the trained model can successfully predict outcomes across different policy implementations.

\section{METHODOLOGY}

Our objective is to learn a function $F$ that maps a set of input simulation parameters $P_{sim}$ to the corresponding simulation output $S$:
\begin{equation}
    F(P_{sim}) \rightarrow S
\end{equation}

In our case, the simulation models the impact of an intervention on a transportation network. The output \( S \) consists of changes in car volume across different edges in the network. More specifically, for each edge \( e \in E \), we aim to predict the change in car volume \( y_e \) due to the intervention. We define \( y_e \) as the simulated change and \( \hat{y}_e \) as the corresponding prediction made by our surrogate model. This defines a standard regression problem, where we learn a function \( F: \mathbb{R}^{|E|\times n} \to \mathbb{R}^{|E|} \) that maps $n$ road characteristics per edge to the predicted change due to an intervention. 

To train the model, we use the Mean Squared Error (MSE) loss function, which is standard for regression tasks. It quantifies the difference between actual and predicted values:

\begin{equation}
\label{eq:loss_function}
\mathcal{L}(F) = \frac{1}{|E|} \sum_{e \in E} (F(x_e) - y_e)^2,
\end{equation}
where $x_e$ denote the road characteristics for road $e$. The objective is to find the optimal function \( F^* \) that minimizes this loss:

\begin{equation}
F^* = \arg \min_F \mathcal{L}(F).
\label{eq:min_F}
\end{equation}

The model utilizes a graph representation in which nodes are characterized by \textit{static}, \textit{positional}, and \textit{variable} features linked to the roads of the street network. Because most GNNs are optimized for node-based predictions, we convert the simulation outputs into street network graphs and then transform them into dual graphs, where nodes correspond to road segments and edges capture the connections between segments. (1) \textit{Static features} provide fixed attributes for each street, such as car volume and capacity in the base case, speed limit and segment length.  (2) \textit{Positional features} capture the spatial information of each street, including the coordinates of its start- and endpoint. (3) \textit{Variable features} represent attributes modified by implemented policies, such as reductions in speed or capacity, or discounts in public transport fees. 

The \textit{base case} refers to a scenario without any policy intervention. It can be created by running the simulation tool multiple times with different random seeds and averaging the features on node (i.e., street) level.

\begin{figure}
    \centering
    \includegraphics[width=\textwidth]{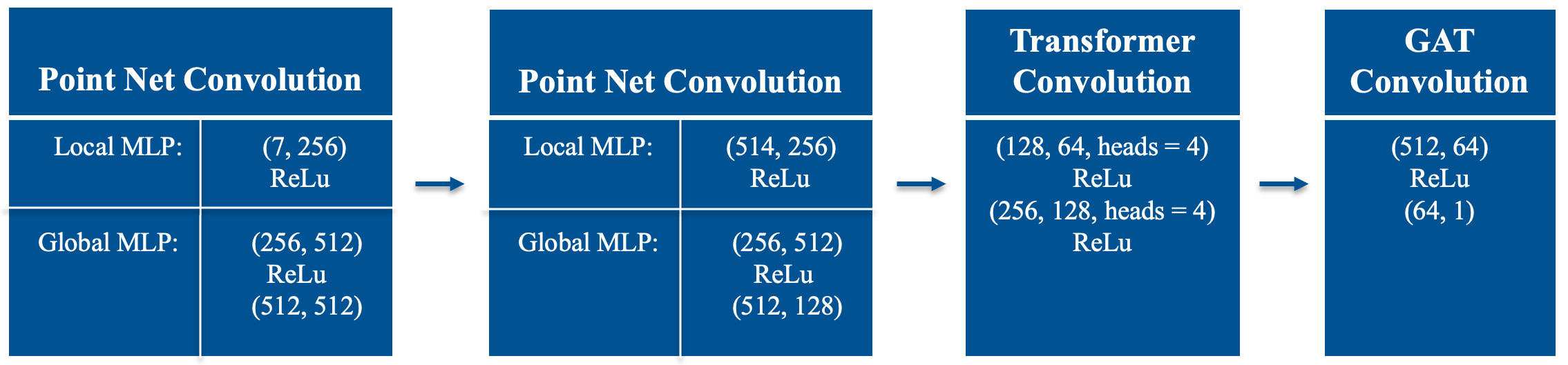}
    \caption{Architecture of the Graph Neural Network}
    \label{fig:gnn_architecture}
\end{figure}

The output predicts the change in car volume on each node (corresponding to a street in the road network) due to the policy in place. The architecture of the GNN, shown in Figure \ref{fig:gnn_architecture}, begins with two layers of PointNet Convolutions (\cite{point_net_2017_2}), which incorporate spatial features and process the positional attributes of each edge. This is followed by two layers of Transformer Convolutions, and one GAT Convolution layer completing the model.

In addition to evaluating the model using the Mean Squared Error (MSE), as defined in \ref{eq:loss_function}, we assess its performance with the coefficient of determination (\( R^2 \)), offering a complementary measure of effectiveness. To establish a performance baseline, we define the \textit{Naïve} MSE as the mean change in car volume across all nodes resulting from the policy.

\section{RESULTS}

We evaluate the proposed methodology using the MATSim model for Paris, France, based on a 1\% downsampled population and 10,000 scenarios. 
We generate the data by first generating district combinations of its 20 districts (``Arrondissements'') by randomly sampling the districts so that the set length follows a normal distribution with a mean of $5$, thereby preventing bias toward larger combinations. The tested policy involves a 50\% capacity reduction on roads classified by OpenStreetMap as ``Primary'', ``Secondary'', or ``Tertiary'' within selected district combinations. Consequently, the model’s variable features are the reduction in capacity. To establish the base case, we run MATSim without policy interventions using 50 random seeds. We then simulate these scenarios to obtain the \textit{ground truth} $y$. Finally, we split the resulting dataset into training, validation and test set with ratios of 80\%, 15\%, 5\%.

The trained model is able to predict the effect of this policy interventions in random combination of districts within 0.1 seconds. It has an accuracy of 0.76 $R^2$ and a validation loss (MSE) of 24.95 (baseline: 103.47) on the test set. Table \ref{tab:result_table} presents the evaluation metrics for the model's performance across various road types. 


\begin{table}
    \centering
    \caption{Result table for different road types}
    \label{tab:result_table}
    \begin{tabular}{lccc} \hline
    Road subset & $R^2$ & Naïve MSE  & Predicted MSE \\ \hline
    All roads & 0.76 & 103.47 & 24.95 \\
    Trunk roads & 0.51 & 216.81 & 105.28 \\
    Primary roads & 0.86 & 407.84 & 59.11 \\
    Secondary roads & 0.65 & 82.67 & 29.10 \\
    Tertiary roads & 0.58 & 47.84 & 19.87 \\
    Roads with policy in place & \textbf{0.92} & 600.95 & 45.31 \\
    Roads without policy in place & 0.39 & \textbf{36.63} & \textbf{22.37} \\ \hline
    \end{tabular}
\end{table}

\begin{figure}
    \centering
    \begin{subfigure}[b]{0.48\textwidth}
        \centering
        \includegraphics[width=\textwidth]{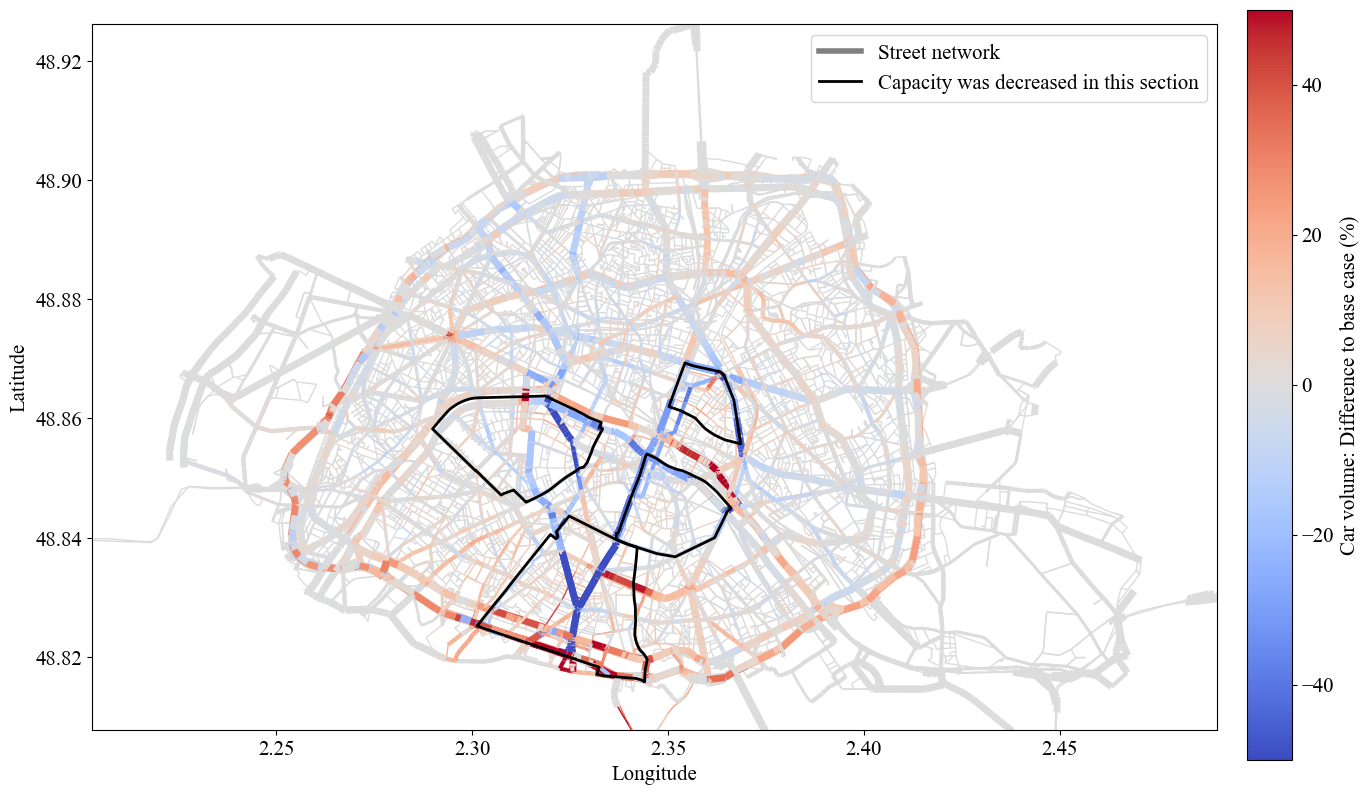}
    \end{subfigure}
    \hfill
    \begin{subfigure}[b]{0.48\textwidth}
        \centering 
        \includegraphics[width=\textwidth]{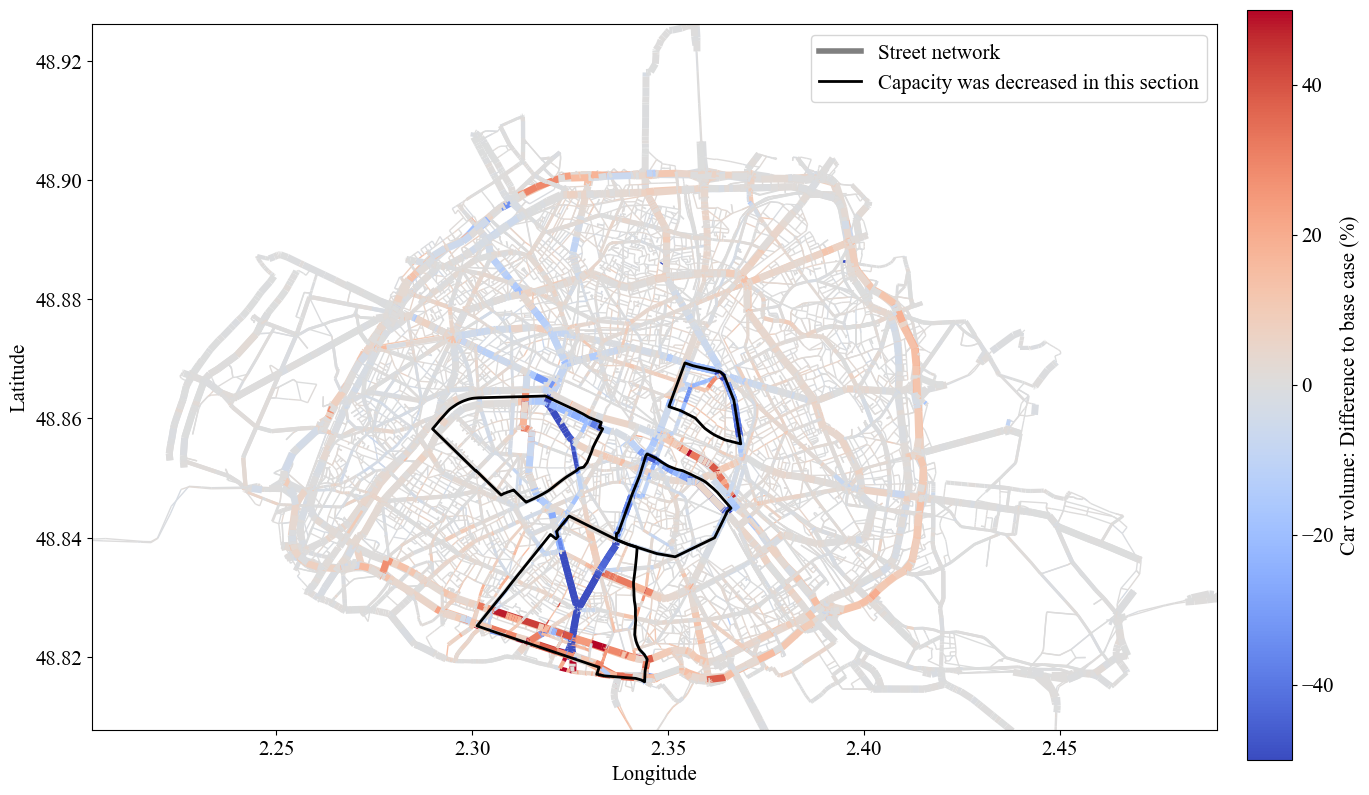}
    \end{subfigure}
    \caption{Randomly chosen test data: Actual (left) and predicted (right) change in car volume}
    \label{fig:result_figures}
\end{figure}

The model demonstrates its highest performance in terms of $R^2$, with a score of $0.92$, on roads where a policy has been implemented. This suggests that the model effectively learned the relationship between capacity reduction and decreased car volume.
Primary roads show the model’s second-best performance in terms of $R^2$, likely due to the high traffic volume, which magnifies the effects of applied policies, allowing for more accurate predictions with reduced stochastic variation. This high traffic volume may also explain the elevated MSE baseline of $407.84$: The mean car volume proves inadequate for primary roads, where usually greater traffic volume is observed.
The lowest MSE appears on roads without policy intervention, with a baseline MSE of \(36.63\) and a predicted MSE of \(22.37\). This indicates that the average car volume after intervention is a good predictor for these roads, suggesting minimal impact from policies applied elsewhere in the network. This may also account for the model's poorest \(R^2\) performance (\(0.39\)). These findings suggest that the model effectively learns ``direct'' correlations, but can still improve in understanding the indirect effects of policies on non-intervention roads, such as evasive driving behavior. This observation is mirrored in Figure \ref{fig:result_figures}, illustrating the predicted versus actual changes in car volume at the street level for a randomly selected district combination from the test set. The model effectively captures the reduction in car volume, highlighted in varying shades of blue, indicating decreases of up to -50\%; Driver evasive behavior, shown in shades of red representing increases of up to 50\%, is slightly less pronounced. For this scenario, the model achieves a test loss of 3.66 (baseline is 137.67) and $R^2$ score of 0.81.

\section{DISCUSSION}

This paper presents a first approach of using Graph Neural Networks as surrogates for large-scale agent-based simulation models. In a case study using the MATSim model of Paris, the GNN successfully learned the impact of capacity reduction policies on the city wide traffic flow.

In the long run, this model could enable a range of applications due to the GNN’s rapid evaluation capabilities. Firstly, simulation-based optimization for determining optimal policy setups becomes feasible, allowing efficient exploration of large solution spaces. Secondly, real-time control applications become possible, as the model enables swift evaluation of scenarios when timely decision-making is crucial.

Achieving this vision requires the following future steps: 1) Expanding predictions to include multimodal impacts, such as mode shifts and public transit usage. 2) Testing a variety of policies to assess broader applicability. 3) Incorporating dynamic scenarios to enable real-time control applications. 4) Developing methods to transfer the trained model across different cities, enhancing its general applicability.

\section{Acknowledgments}
We thank the German Federal Ministry of Transport and Digital Infrastructure for providing funding through the project ``MINGA'' with grant number 45AOV1001K. We remain responsible for all findings and opinions presented in the paper.

We thank Alejandro Tejada Lapuerta from Helmholtz Munich for fruitful discussions about setting up the model and interpreting the results.

Chat-GBT 4.0 was used for summarizing paragraphs. 

\section{Author contributions}

The authors confirm their contributions to the paper as follows: E. Natterer, R. Engelhardt, S. Hörl, and K. Bogenberger conceived and designed the study. S. Hörl provided the MATSim simulation for Paris. E. Natterer executed the MATSim simulations and developed the machine learning model. E. Natterer and R. Engelhardt analyzed and interpreted the results. All authors reviewed the results and approved the final version of the manuscript.



\bibliographystyle{trb}
\bibliography{references,sebastians_bib}

\end{document}